\documentclass[a4paper]{jpconf}

\usepackage{graphicx}
\usepackage{color}

\usepackage{cite,epsfig}
 
 \usepackage{amsmath,amssymb}

\usepackage{ifthen,array}
 \usepackage{setspace}
 \usepackage{fancyhdr}
 \usepackage{moreverb}
 \usepackage{rotating}
\usepackage{amsfonts}
\usepackage{bbm}

\newcommand{\Atm}  {\textsc{atm}}

\newcommand{\Dms}  {\Delta m^2_{21}}
\newcommand{\Dma}  {\Delta m^2_{31}}

\def\e6{$E(6)$}
\def\10{$SO(10)$}
\def\21{$SU(2) \otimes U(1) $}

\def\422{$SU(4) \otimes SU(2) \otimes SU(2)$}
\def\321{$SU(3) \otimes SU(2) \otimes U(1)$}
\def\lsim{\raise0.3ex\hbox{$\;<$\kern-0.75em\raise-1.1ex\hbox{$\sim\;$}}}
\def\gsim{\raise0.3ex\hbox{$\;>$\kern-0.75em\raise-1.1ex\hbox{$\sim\;$}}}
\def\lfv{lepton flavour violation }

\def\meff{\langle m_{\nu} \rangle}
\newcommand{\ed}{\end{document}}
\DeclareMathAlphabet{\mathsc}{OT1}{cmr}{m}{sc}

\newcommand{\eps}{\epsilon_{13}}

\def \nbb {$\beta\beta_{0\nu}$ }

\def\meff{\langle m_{\nu} \rangle}

\let\vev\VEV

\def\e6{$E(6)$}
\def\10{$SO(10)$}
\def\21{$SU(2) \otimes U(1) $}

\def\422{$SU(4) \otimes SU(2) \otimes SU(2)$ }
\def\321{$SU(3) \otimes SU(2) \otimes U(1)$ }

\newcommand{\AddrAHEP}{%
 AHEP Group, Instituto de F\'{\i}sica Corpuscular,
  C.S.I.C. -- Universitat de Val{\`e}ncia \\
  Edificio de Institutos de Paterna, Apartado 22085,
  E--46071 Val{\`e}ncia, Spain\\}
\begin{document}
\title{Progress in the understanding of neutrino properties}

\author{J. W. F. Valle}

\address{\AddrAHEP}

\ead{valle@ific.uv.es}

\begin{abstract}
  I briefly summarize neutrino oscillation results and discuss their
  robustness. I mention recent attempts to understand the pattern of
  neutrino mixing within various seesaw mechanisms, with or without
  supersymmetry and/or flavor symmetries.
  I also mention the possibility of intrinsic supersymmetric neutrino
  masses in the context of broken R parity models, showing how this
  leads to clear tests at the LHC.
\end{abstract}

\section{Neutrino Oscillations}
\label{sec:neutr-oscill}
\vglue .1cm

\begin{figure}[!h]
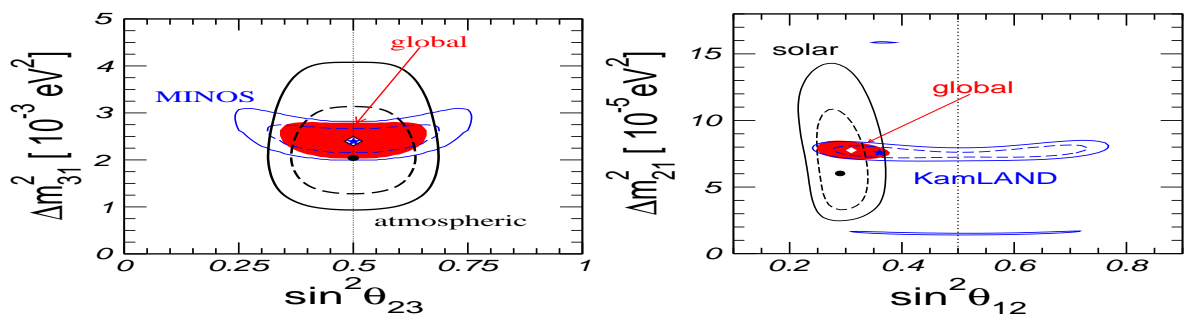
 \centering
\includegraphics[width=.48\linewidth,height=4cm]{atm-new.eps}
\includegraphics[width=.48\linewidth,height=4cm]{sol-new.eps}
\caption{\label{fig:global} %
Neutrino oscillation parameters from a global analysis of the world's
current data~\cite{Schwetz:2008er}. }
\end{figure}

\begin{figure}[h]
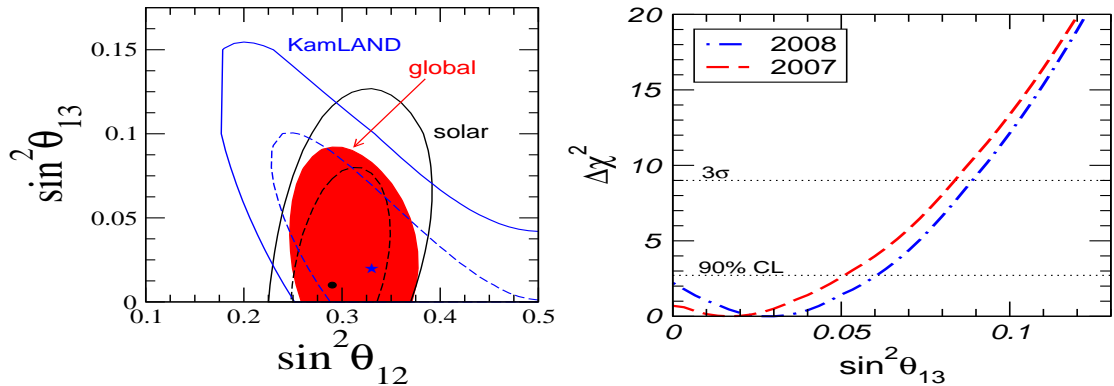
 \centering
\includegraphics[height=5cm,width=.45\linewidth]{s12-s13-tension.eps}
\includegraphics[height=5cm,width=.45\linewidth]{th13-sol-07vs08-lin.eps}
\caption{\label{fig:th13}%
Constraints on $\sin^2\theta_{13}$ from different parts of
  the global data given in Ref.~\cite{Schwetz:2008er}.}
\end{figure}

Neutrino flavors inter-convert during propagation both \emph{in vacuo}
and in matter.  Uncontroversial evidence for this follows from solar
and atmospheric, as well as reactor and accelerator neutrino
studies~\cite{exp-talks-taup09}.
The basic concept needed to describe oscillations is the lepton mixing
matrix, the leptonic analogue of the quark mixing matrix. In its
simplest $3\times 3$ unitary form it is given
as~\cite{schechter:1980gr}
\begin{equation}
  \label{eq:2227}
K =  \omega_{23} \omega_{13} \omega_{12}
\end{equation}
where each factor is effectively $2\times 2$ with an angle and a
corresponding CP phase. Since current experiments are insensitive to
CP violation, we neglect phases, so that oscillations depend only on
the three mixing angles $\theta_{12}, \theta_{23}, \theta_{13}$ and on
the two squared-mass splittings $\Dms \equiv m^2_2 - m^2_1$ and $\Dma
\equiv m^2_3 - m^2_1$ associated to solar and atmospheric
transitions~\cite{Maltoni:2004ei}.
To a good approximation, one can also set $\Dms = 0$ in the analysis
of atmospheric and accelerator data, and $\Dma$ to infinity in the
analysis of solar and reactor data.  The resulting neutrino
oscillation parameters are summarized in Figs.~\ref{fig:global} and
\ref{fig:th13}, taken from~\cite{Schwetz:2008er}. \vskip .001cm
The left and right panels in Fig.~\ref{fig:global} give the
``atmospheric'' and ``solar'' oscillation parameters, $\theta_{23}$ \&
$\Dma$, and $\theta_{12}$ \& $\Dms$, respectively, minimizing with
respect to the undisplayed parameters in each case, and including
always all relevant data.  The dot, star and diamond indicate the best
fit points of atmospheric MINOS and global data, as well as solar,
KamLAND and global data, respectively.
Note the complementarity between data from artificial and natural
neutrino sources: reactors and accelerators give the best
determination of squared-mass-splittings, while mixings are mainly
determined by solar and atmospheric data.
Current data slightly prefer a nonzero value of the remaining angle
$\theta_{13}$, as seen in Fig.~\ref{fig:th13}. Since the hint is not
yet significant, we prefer to interpret this as a weaker bound on
$\theta_{13}$~\footnote{Note: the
  bounds in Eq.~(\ref{eq:th13a}) are given for 1~dof, while the
  regions in Fig.~\ref{fig:th13} (left) are 90\%~CL for 2~dof}:

\begin{equation}\label{eq:th13a}
  \sin^2\theta_{13} \le \left\lbrace \begin{array}{l@{\qquad}l}
      0.060~(0.089) & \text{(solar+KamLAND)} \\
      0.027~(0.058) & \text{(CHOOZ+atm+K2K+MINOS)} \\
      0.035~(0.056) & \text{(global data)}
    \end{array} \right.
\end{equation}
If confirmed by future data, a nonzero $\theta_{13}$ would encourage
the search for CP violation in upcoming neutrino oscillation
experiments~\cite{Bandyopadhyay:2007kx,Nunokawa:2007qh}.
The expected CP asymmetries are small, as they are suppressed both by
$\theta_{13}$ and by the small ratio $ \alpha$ of solar/atmospheric
squared mass splittings, currently determined as $ \alpha = 0.032\,,
\quad 0.027 \le \alpha \le 0.038 \quad (3\sigma) \,.  $

\section{Are oscillations robust?} 
 \label{sec:robustn-oscill}
\vglue .1cm

Many effects may distort the ``celestial'' neutrino fluxes reaching
the detectors, and hence affect the determination of oscillation
parameters. In this connection several
regular~\cite{miranda:2000bi,guzzo:2001mi,barranco:2002te} and
random~\cite{Miranda:2003yh,Miranda:2004nz} convective zone solar
magnetic field models have been considered.
 These fields would induce neutrino spin-flavor
 precession~\cite{schechter:1981hw,akhmedov:1988uk,Lim:1987tk} and
 hence modulate the originally produced standard solar model neutrino
 fluxes.
 Similarly, the presence of radiative zone random magnetic fields
 could also induce density fluctuations deep inside the
 Sun~\cite{Burgess:2003fj} and substantially modify the
 energy-dependence of the solar neutrino survival
 probability~\cite{Loreti:1994ry,nunokawa:1996qu}, with a potentially
 important impact on the determination of oscillation parameters.

 Reactor neutrino data from KamLAND have played a crucial role in
 probing the robustness of solar oscillations against astrophysical
 uncertainties, such as radiative~\cite{burgess:2002we,Burgess:2003su}
 or convective zone~\cite{miranda:2000bi,guzzo:2001mi,barranco:2002te}
 magnetic fields. The stringent limits on solar anti-neutrinos from
 KamLAND lead to strong limits on neutrino magnetic transition
 moments, especially in the case of turbulent
 fields~\cite{Miranda:2003yh,Miranda:2004nz}.
 The result is that oscillations constitute the only viable
 explanation of the data, and of the oscillation solutions allowed by
 solar data~\cite{Gonzalez-garcia:2000sq}, only the large mixing angle
 solution is consistent with the spectrum measurements at
 KamLAND~\cite{pakvasa:2003zv}.

\begin{figure}[!h] \centering
 \includegraphics[height=2.7cm,width=.4\linewidth]{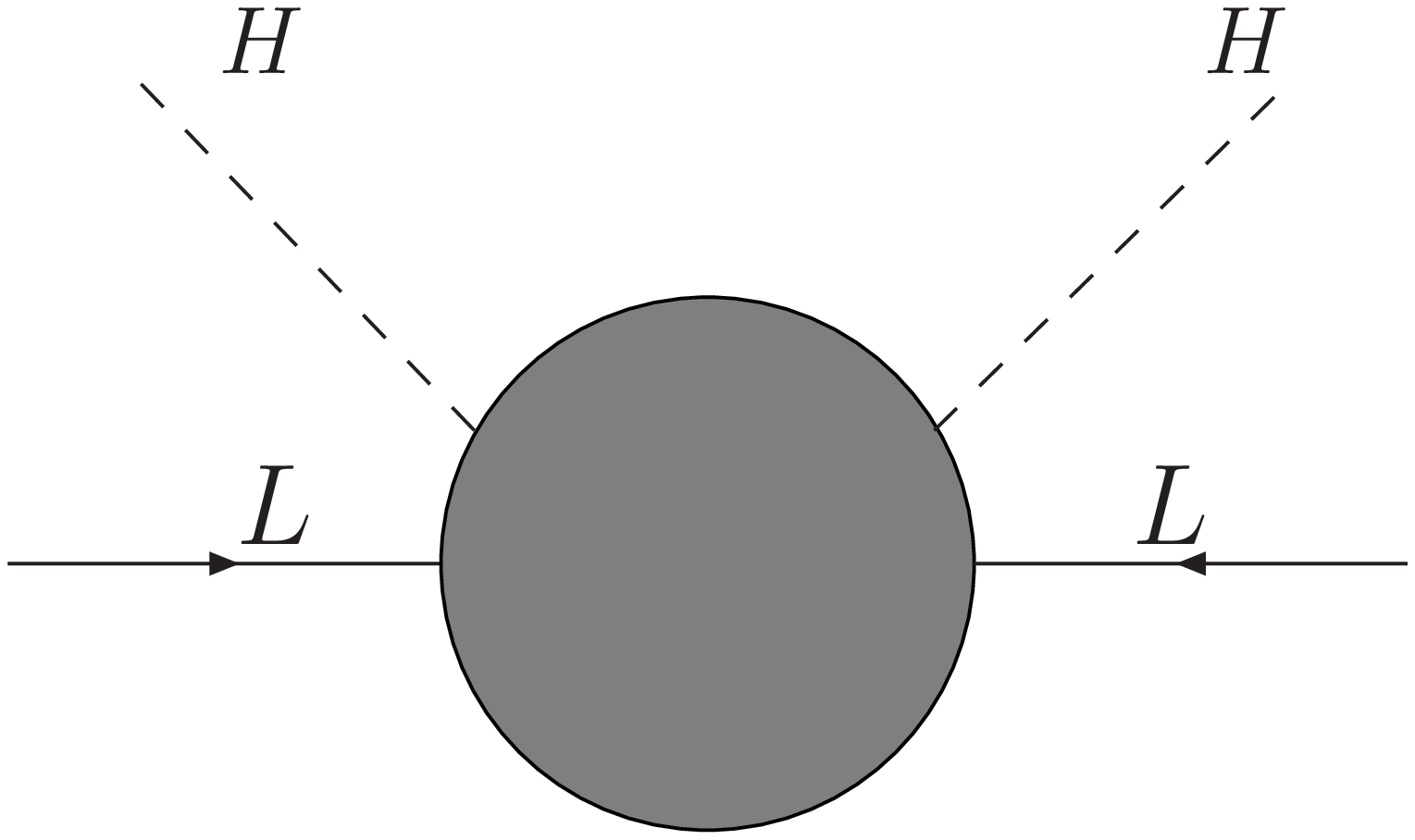}
\hskip 1cm
 \includegraphics[height=2.7cm,width=.4\linewidth]{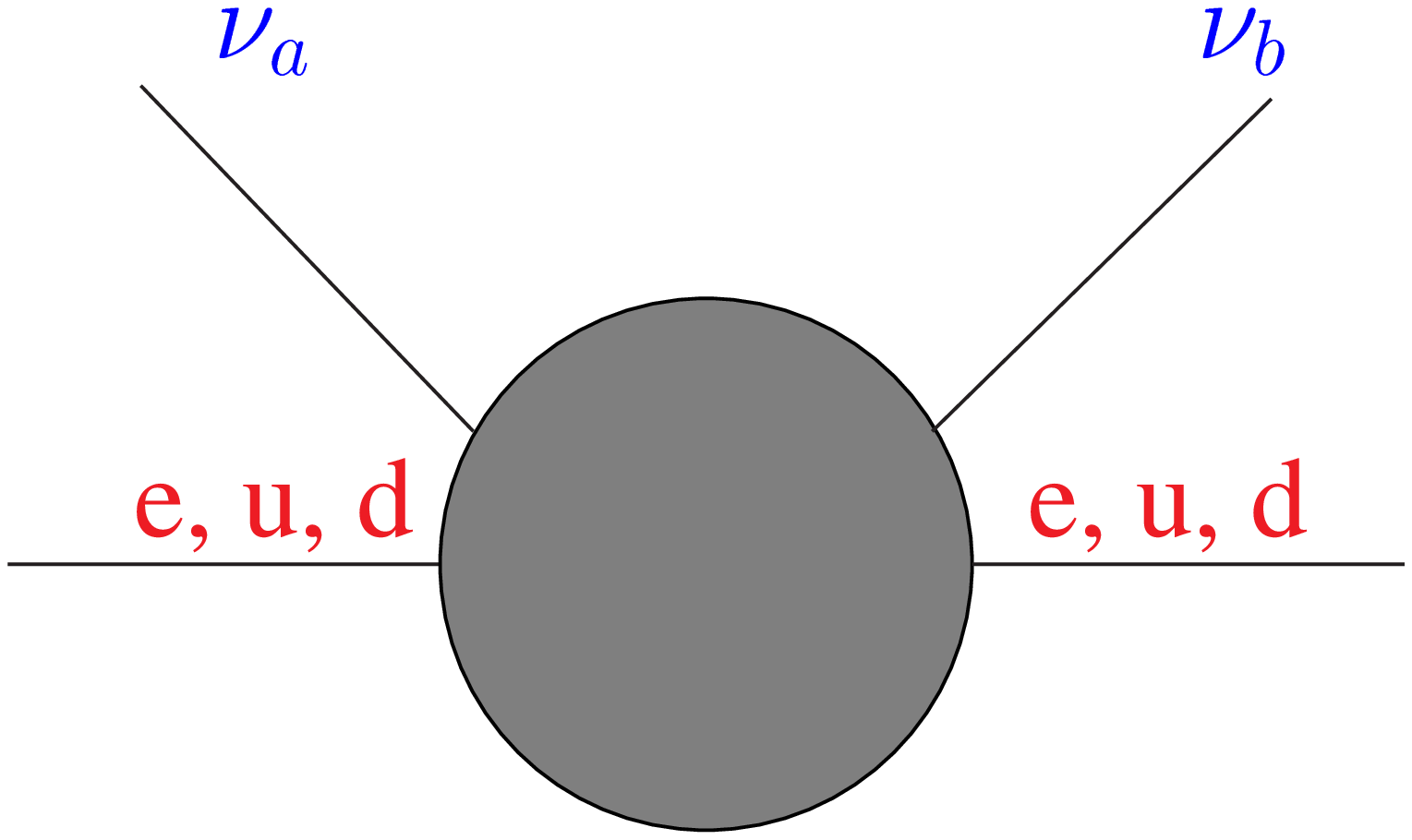}
\caption{\label{fig:d-5-nsi} %
    Neutrino mass and corresponding NSI operators.}
\end{figure}

 In Standard Model language neutrino mass is described by an effective
 dimension-five operator, shown in the left panel in
 Fig.~\ref{fig:d-5-nsi}.\\ \vskip -.4cm

Most neutrino mass generation schemes also induce sub-weak strength
($\sim \varepsilon G_F$) dimension-6 operators depicted in the right
panel in Fig.~\ref{fig:d-5-nsi}. They can be of two types:
flavour-changing (FC) and non-universal (NU).
For example, the presence of NSI is expected low-scale seesaw models,
such as the
inverse~\cite{mohapatra:1986bd,Bazzocchi:2009kc,Ibanez:2009du} and the
linear~\cite{Malinsky:2005bi} seesaw models, where the non-unitary
piece~\cite{schechter:1980gr} of the lepton mixing matrix can be
sizeable, hence the induced non-standard interactions.  Relatively
sizable NSI strengths may also be induced in models with radiatively
induced neutrino
masses~\cite{zee:1980ai,babu:1988ki,AristizabalSierra:2006ri,AristizabalSierra:2006gb}. The
strength of the NSI operators will play an important role in
elucidating the origin of neutrino mass, as it will help discriminate
between high and low-scale schemes.

\begin{figure*}
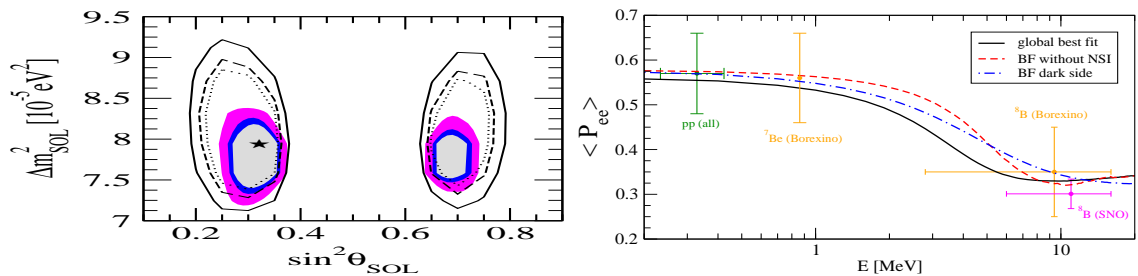

\begin{center}
 \includegraphics[height=3.5cm,width=.46\textwidth,angle=0]{sol-panel.eps}
 \includegraphics[height=3.5cm,width=.46\textwidth,angle=0]{Pee-profile.eps}
 \caption{Left: normal and ``dark-side'' solutions in the presence of
   non-standard neutrino interactions (NSI). Right: solar neutrino
   survival probabilities averaged over the $^8$B neutrino production
   region for three reference points, see
   Ref.~\cite{Escrihuela:2009up} for details. Vertical bars correspond
   to experimental errors, while horizontal ones indicate the energy
   range relevant for each experiment. }
\label{fig:nsi-pee}
\end{center}
\end{figure*}

As noted in Ref.~\cite{Miranda:2004nb}, current determination of solar
neutrino parameters is not yet robust against the existence of
non-standard neutrino interactions. In fact, in the presence of such
interactions, there is a new ``dark side''
solution~\cite{Miranda:2004nb}, as shown in the left panel in
Fig.~\ref{fig:nsi-pee}, taken from Ref.~\cite{Escrihuela:2009up}. This
solution is almost degenerate with the usual one, and survives the
inclusion of reactor data.
As seen in the right panel in Fig.~\ref{fig:nsi-pee}, the lines are
compared with the experimental rates for the pp neutrino flux (from
the combination of all solar experiments), the 0.862 MeV $^7$Be line
(from Borexino) and two estimated values of the $^8$B neutrino flux
from Borexino and SNO (third phase). Distinguishing between these
solutions poses a challenge for low energy solar neutrino experiments.

As already noted, the slight tension between current solar and KamLAND
data is alleviated allowing for a non-zero value of the mixing angle
$\theta_{13}$. Likewise, non-standard flavor-changing interactions may
also alleviate this tension and hence constitute a source of confusion
in the determination of $\theta_{13}$. Indeed, a full three-flavor
analysis of solar and KamLAND data in the presence of flavor-changing
NSI~\cite{Palazzo:2009rb} reveals a degeneracy between $\theta_{13}$
and the vectorial coupling $\eps$ characterizing the non-standard
transitions between $\nu_e$ and $\nu_\tau$ in the forward scattering
process with d-type quarks. 

Non-standard neutrino interactions may also be probed at high
energies.  One can show that already even a small residual NSI
strengths may have dramatic consequences for the sensitivity to
$\theta_{13}$ at a neutrino factory~\cite{huber:2001de,huber:2002bi}.
Improving the sensitivities on NSI constitutes a necessary step and
opens a window of opportunity for neutrino physics in the precision
age.  The presence of NSI also leads to novel effects in supernova
neutrino propagation~\cite{dighe}, for example, the possibility of
internal resonant neutrino conversions even in the absence of neutrino
masses~\cite{valle:1987gv,nunokawa:1996tg,EstebanPretel:2007yu}.

In contrast to the ``solar'' sector, thanks to the large statistics of
atmospheric data over a wide energy range, the determination of $\Dma$
and $\sin^2\theta_\Atm$ is hardly affected by the presence of NSI, at
least within the 2--neutrino
approximation~\cite{fornengo:2001pm}. Future neutrino factories will
substantially improve this bound~\cite{huber:2001zw}.
                                        
\section{How do neutrinos get mass?}
\label{sec:neutrino-mass}
\vglue .1cm

Neutrino oscillations provide the first sign of physics beyond the
Standard Model (SM). 
Pinning-down the ultimate origin of neutrino mass remains a
challenge. Apart from any detailed mass generation mechanism, since
they carry no electric charge they should have Majorana-type
masses~\cite{schechter:1980gr}.  This is indeed what happens in
specific neutrino mass generation schemes, markedly different from
those of charged fermions. The latter come in two chiral species and
get mass linearly in the electroweak symmetry breaking vacuum
expectation value (vev) $\vev{\Phi}$ of the Higgs scalar doublet.  In
contrast, as illustrated in the left panel in Fig.~\ref{fig:d-5-nsi},
neutrino masses arise as an effective lepton number violating
dimension-five~\footnote{Lepton number violation may also show up at
  higher dimension~\cite{Gogoladze:2008wz,Bonnet:2009ej}.}  operator
${\cal O} \equiv \lambda L \Phi L \Phi$ (where $L$ denotes a lepton
doublet)~\cite{Weinberg:1980bf}.
It follows that a simple way to account for the smallness of neutrino
masses, is that the coefficient characterizing the strength of ${\cal
  O}$ is suppressed either by a high-scale $M_X$ in the denominator
or, alternatively, it may involve a low-mass-scale in the numerator.
The search of NSI will help discriminate which of the two pathways is
chosen by nature.

Gravity is often argued to break global
symmetries~\cite{Coleman:1988tj,Kallosh:1995hi}, and would induce
Weinberg's operator ${\cal O}$, with $M_X$ identified to the Planck
scale $M_P$. The resulting masses are too small, implying the need for
physics beyond the SM, typically at a high scale below
$M_P$~\cite{deGouvea:2000jp}.
Alternatively ${\cal O}$ may be suppressed by small scales, Yukawa
couplings and/or loop-factors~\cite{Valle:2006vb}. This way one may
have tree level, radiative, and hybrid mechanisms, all of which may
have high- or low-scale realizations.
Depending on whether it is gauged or not, spontaneous lepton-number
violation implies either an extra neutral gauge boson or a
Nambu-Goldstone boson coupled to neutrinos.
One may construct models of both types.
However the most basic and general seesaw description is in terms of
the \321 gauge structure~\cite{schechter:1980gr}. In such a framework
the relevant scale can either be large or small, depending on model
details, with a fair chance that the origin of neutrino mass may be
probed at accelerators like the LHC.

\vglue -0.5cm
\subsection{Minimal seesaw }
\label{sec:i-minimal-seesaw}
\vglue .1cm

\begin{figure}[h] 
\centering
 \includegraphics[scale=.12,width=.45\linewidth]{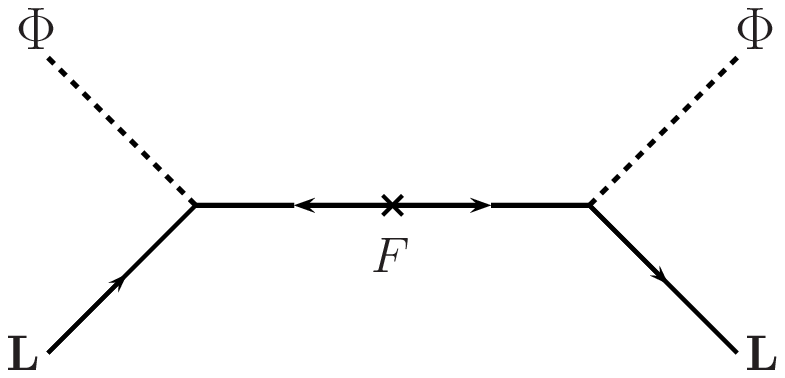} 
 \includegraphics[height=2cm,width=.45\linewidth]{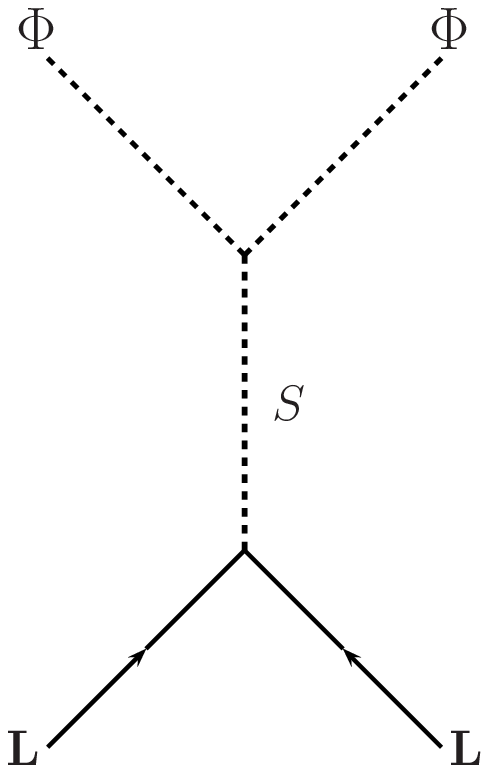}
 \caption{\label{fig:seesaw} Type-I and III~(left) and Type-II~(right)
   realizations of the seesaw mechanism.}\vglue -.2cm
 \end{figure}

Here the operator ${\cal O}$ is induced by the exchange of heavy
states with masses close to the ``unification'' scale, as indicated in
Fig.~\ref{fig:seesaw}, for a review see Ref.~\cite{Valle:2006vb}.
In type-I or type-III seesaw the exchanged states are fermions
singlets or triplets, while {Type-II} seesaw is induced by triplet
scalar exchange. 
 The ``complete seesaw'' has both fermion and scalar intermediate
 states and its phenomenology has been thoroughly studied
 in~\cite{schechter:1980gr} including the general seesaw
 diagonalization method~\cite{schechter:1982cv}. The latter is
 obtained using the hierarchy of vevs $v_3 \ll v_2 \ll v_1$ needed in
 order to account naturally for the small neutrino masses and which
 follows by minimizing the scalar potential. 

\subsection{``Non-minimal'' seesaw}
\label{sec:ii-non-missionaire}
\vglue .1cm

The seesaw is not a model but a general paradigm. One may implement it
with different gauge groups and multiplet contents, with gauged or
ungauged B-L, broken explicitly or spontaneously, at a high or at a
low energy scale, with or without
supersymmetry. 
An attractive class of extended seesaw schemes employs extra gauge
singlet neutrinos in addition to the $\nu^c$ which are present in the
16 of \10 group~\cite{mohapatra:1986bd}.
One may implement such schemes with low-scale breaking of
B-L~\cite{Malinsky:2005bi,Hirsch:2006ft}, leaving a TeV-scale
$Z^\prime$~\cite{valle:1987sq}.  However, whatever the symmetry is, it
must ultimately break to \321, hence the most general seesaw
description must be formulated at this
level~\cite{schechter:1980gr}. Such low-energy description is crucial
in accurately describing low-scale variants of the seesaw mechanism,
where the basic lepton-number-violating parameter may be naturally
small and calculable due to supersymmetric renormalization group
evolution~\cite{Bazzocchi:2009kc}.

\vglue -0.5cm

\subsection{Radiative schemes}
\label{sec:ii-radiative-schemes}
\vglue .1cm 

The operator ${\cal O}$ may arise as loop
effects~\cite{zee:1980ai,babu:1988ki}, with no need for a large scale.
In this case its coefficient $\lambda$ is suppressed by small
loop-factors, Yukawa couplings and possibly by a small scale parameter
characterizing the breaking of lepton number, leading to naturally
small neutrino masses.
Like low-scale seesaw models discussed above, radiative schemes
typically have new TeV states opening the door to phenomenology at
the LHC.~\cite{AristizabalSierra:2007nf}.

\vglue -0.5cm

\subsection{R parity violation}
\label{sec:vvv rpv}
\vglue .1cm The origin of neutrino masses may be intrinsically
supersymmetric in models where the R parity symmetry
breaks~\cite{hall:1984id,Ross:1984yg,Ellis:1984gi}.
This may happen spontaneously, driven by a nonzero vev of an \321
singlet sneutrino~\cite{Masiero:1990uj,romao:1992vu,romao:1997xf},
leading to effective bilinear R-parity
violation~\cite{Diaz:1998xc,Hirsch:2004he}.
This provides the minimal way to break R parity in the Minimal
Supersymmetric Standard Model~\cite{Hirsch:2004he}. The induced
neutrino mass spectrum is hybrid, with one scale (typically the
atmospheric) induced by neutralino-exchange at the tree level, and the
other scale (solar) induced by \emph{calculable} one-loop
corrections~\cite{Hirsch:2000ef}.

Unprotected by any symmetry, the lightest supersymmetric particle
(LSP) decays, typically inside detectors at the Tevatron or the
LHC~\cite{Hirsch:2000ef,Diaz:2003as}.  The LSP decay-length is a
measure of the neutrino mass and can be experimentally resolvable,
leading to a displaced vertex~\cite{deCampos:2005ri,deCampos:2007bn}.
More strikingly, its decay properties correlate with the neutrino
mixing angles. Indeed, the LSP decay pattern is predicted by the
low-energy measurement of the atmospheric angle
$\theta_{23}$~\cite{Porod:2000hv,romao:1999up,mukhopadhyaya:1998xj}.
Such a prediction will be clearly tested at the LHC. 
Similar correlations hold in schemes based on other supersymmetry
breaking mechanisms and, correspondingly, featuring other states as
LSP~\cite{Hirsch:2003fe}.

\vglue -0.5cm
\section{Understanding lepton mixing with flavor symmetries}
\label{sec:Flavor}
\vglue .1cm

As seen above current neutrino oscillation data indicate solar and
atmospheric mixing angles which are unexpectedly large when compared
with quark mixing angles. This places a challenge to the understanding
of the flavor problem in unified schemes where quarks and leptons are
related.
It has been noted that the neutrino mixing angles are approximately
tri-bi-maximal~\cite{Harrison:2002er}.  There have been many schemes
suggested in the literature in order to reproduce this pattern using
various discrete flavor symmetry groups containing mu-tau symmetry,
e.~g.~\cite{babu:2002dz,Hirsch:2003dr,Harrison:2002et,Grimus:2003yn,
  Altarelli:2005yp,Mondragon:2007af,Bazzocchi:2009da,Altarelli:2009gn,Grimus:2009mm,Joshipura:2009tg}.
One expects the flavor symmetry to be valid at high energy scales.
Deviations from the tri-bi-maximal ansatz~\cite{King:2009qt} may be
calculable by renormalization group
evolution~\cite{Antusch:2003kp,Plentinger:2005kx,Hirsch:2006je}.

A specially simple ansatz is that, as a result of a given flavor
symmetry such as A4~\cite{babu:2002dz,Hirsch:2003dr}, neutrino masses
unify at high energies $M_X$~\cite{chankowski:2000fp}, the same way as
gauge couplings do in the presence of
supersymmetry~\cite{Raby:2009sf}. Such quasi-degenerate neutrino
scheme predicts $\theta_{23}=\pi/4~\rm{and}~\theta_{13}=0$, leaving
the solar angle $\theta_{12}$ unpredicted, though
Cabibbo-unsuppressed.  If CP is violated $\theta_{13}$ becomes
arbitrary and the Dirac phase is maximal~\cite{Grimus:2003yn}.  One
can show that lepton and slepton mixings are related and that at least
one slepton lies below 200 GeV, within reach of the LHC. The lower
bound on the absolute Majorana neutrino mass scale $m_0 \gsim 0.3$ eV
ensures that the model will be probed by future cosmological tests and
$\beta\beta_{0\nu}$ searches.  Expected rates for LFV processes
BR$(\mu \to e \gamma) \gsim 10^{-15}$ and BR$(\tau \to \mu \gamma) >
10^{-9}$ typically lie within reach of upcoming experiments.
Note that flavor symmetries, such as our A4, may also be implemented
in low-scale seesaw schemes, both type-I~\cite{Hirsch:2009mx} and
type-III~\cite{Ibanez:2009du}, leading to different neutrino mass
spectra.

\vglue -0.5cm
\section{Lepton flavor violating (LFV) effects}
\label{sec:lfv}
\vglue .1cm

The unequivocal evidence that neutrinos oscillate suggests that, at
some level, flavor violation should also show up as transitions
involving the charged leptons, since these are their electroweak
doublet partners.  There are two basic mechanisms: (i) neutral heavy
lepton
exchange~\cite{Bernabeu:1987gr,gonzalez-garcia:1992be,Ilakovac:1994kj}
and (ii) supersymmetry
exchange~\cite{borzumati:1986qx,casas:2001sr,Antusch:2006vw}.
\begin{figure}[!h]
  \centering
\includegraphics[width=7cm,height=3.3cm]{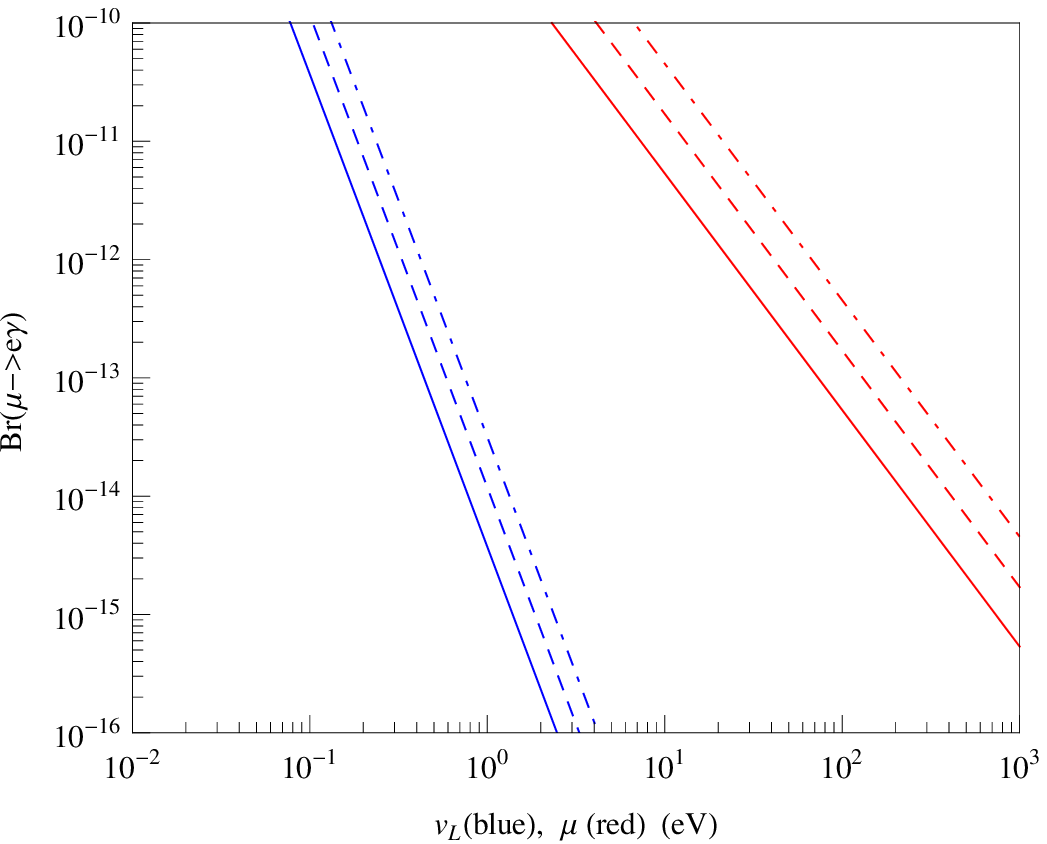}
\includegraphics[width=7cm,height=3.3cm]{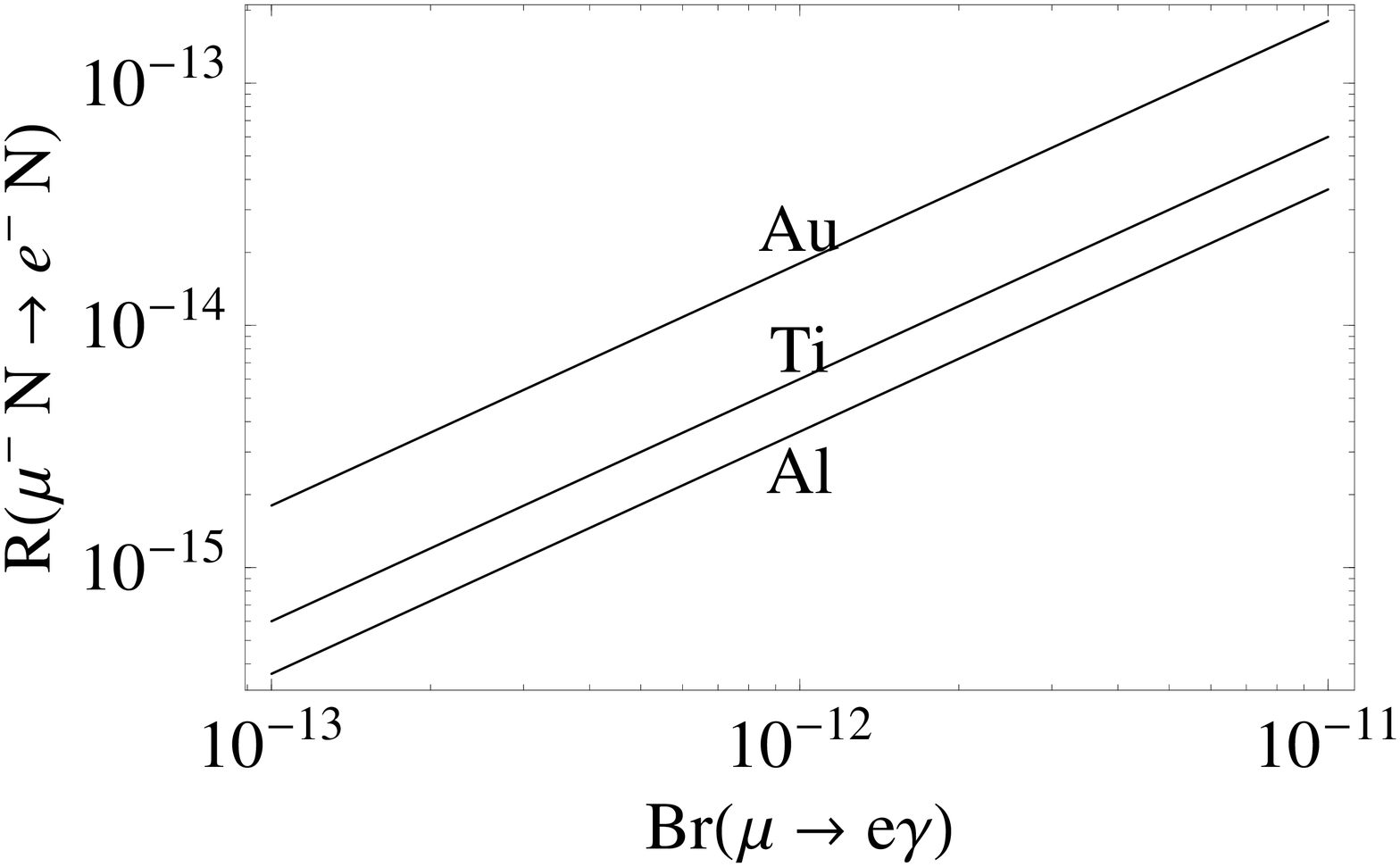}
\caption{Left: $Br(\mu\to e \gamma)$ versus the LNV scale in the
  inverse seesaw in top (red color), and linear seesaw in bottom (blue
  color). The heavy lepton mass $M$ is fixed as $M=100 \, GeV$
  (continuous line), $M=200\, GeV$ (dashed line) and $M=1000\, GeV$
  (dot-dashed line), from~\cite{Hirsch:2009mx}.  Right: typical
  correlation between mu-e conversion in nuclei and \(Br(\mu\to
  e\gamma)\), from~\cite{Deppisch:2005zm}.}
\label{fig:lfv-low-scale} \vglue -.2cm
\end{figure} 
Due to the sizeable admixture of right-handed neutrinos in their
leptonic charged current, low-scale seesaw schemes induce potentially
large LFV rates~\cite{Bernabeu:1987gr,gonzalez-garcia:1992be} and CP
violating processes as well~\cite{branco:1989bn,Rius:1989gk}. This can
happen in the absence of supersymmetry and in the limit of massless
neutrinos, hence their magnitude is unrestricted by the smallness of
neutrino masses.
In Fig.~\ref{fig:lfv-low-scale} we display \(Br(\mu\to e\gamma)\)
versus the small lepton number violating (LNV) parameters $\mu$ and
$v_L$ for two different low-scale seesaw models, the inverse and the
linear seesaw, respectively.  Clearly the LFV rates are sizeable in
both cases. %
Similarly one can see that in low-scale seesaw models the nuclear
$\mu^--e^-$ conversion rates lie within planned sensitivities of
future experiments such as PRISM~\cite{Kuno:2000kd}.
Note that in type-III versions of such low-scale seesaw
schemes~\cite{Ibanez:2009du}, the TeV RH neutrinos would not only
induce LFV processes but also be copiously produced at the LHC.

\begin{figure}[!htb]
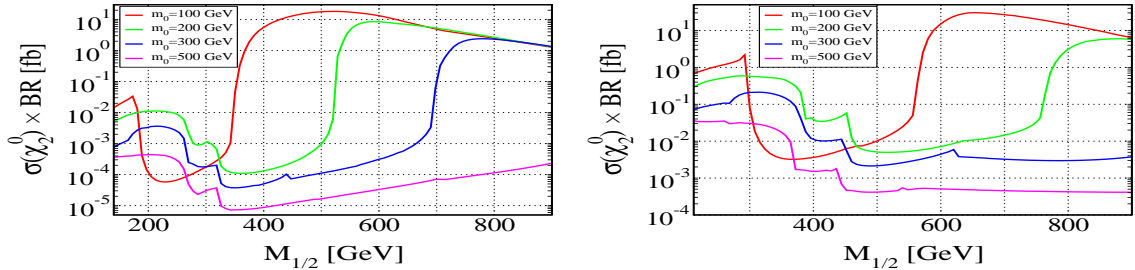

  \centering
\begin{tabular}{cc}
  \includegraphics[width=0.45\textwidth,height=3.5cm]{plot-sigmaLOfbBR-m12_-_4m0.eps} &
  \includegraphics[width=0.45\textwidth,height=3.5cm]{plot-sigmaLOfbBR-m12_-_4m0_-_II.eps}
\end{tabular}
\caption{LFV rate for $\mu$-$\tau$ lepton pair production from
  $\chi^0_2$ decays versus $M_{1/2}$ for the indicated $m_0$ values,
  assuming minimal supergravity parameters: $\mu>0$, $\tan\beta=10$
  and $A_0=0$ GeV, for type-I (left) and for type-II seesaw
  (right). Here $\lambda_1=0.02$ and $\lambda_2=0.5$ are Type-II
  seesaw parameters, and we imposed the constraint Br($\mu\to e
  +\gamma) \le 1.2\cdot 10^{-11}$, from Ref.~\cite{Esteves:2009vg}.}
  \label{fig:ProdXBR} \vglue -.2cm
\end{figure}

In contrast, barring fine-tunings, high-scale seesaw models require
supersymmetry in order to have sizeable LFV rates.  Here \lfv is
expected to show up in the most direct way in the production of
supersymmetric particles at the LHC, as seen in
Fig.~\ref{fig:ProdXBR}.

Both supersymmetric and RH neutrino contributions to \lfv exist in
supersymmetric seesaw schemes, and their interplay depends on the
seesaw scale~\cite{Deppisch:2004fa}.

\vglue -0.5cm
\section{Lepton number violation and  \nbb}
\label{sec:lepton-number-lepton}
\vglue 0.1cm 

\begin{figure}[h]
  \centering
\includegraphics[width=5cm,height=3.5cm]{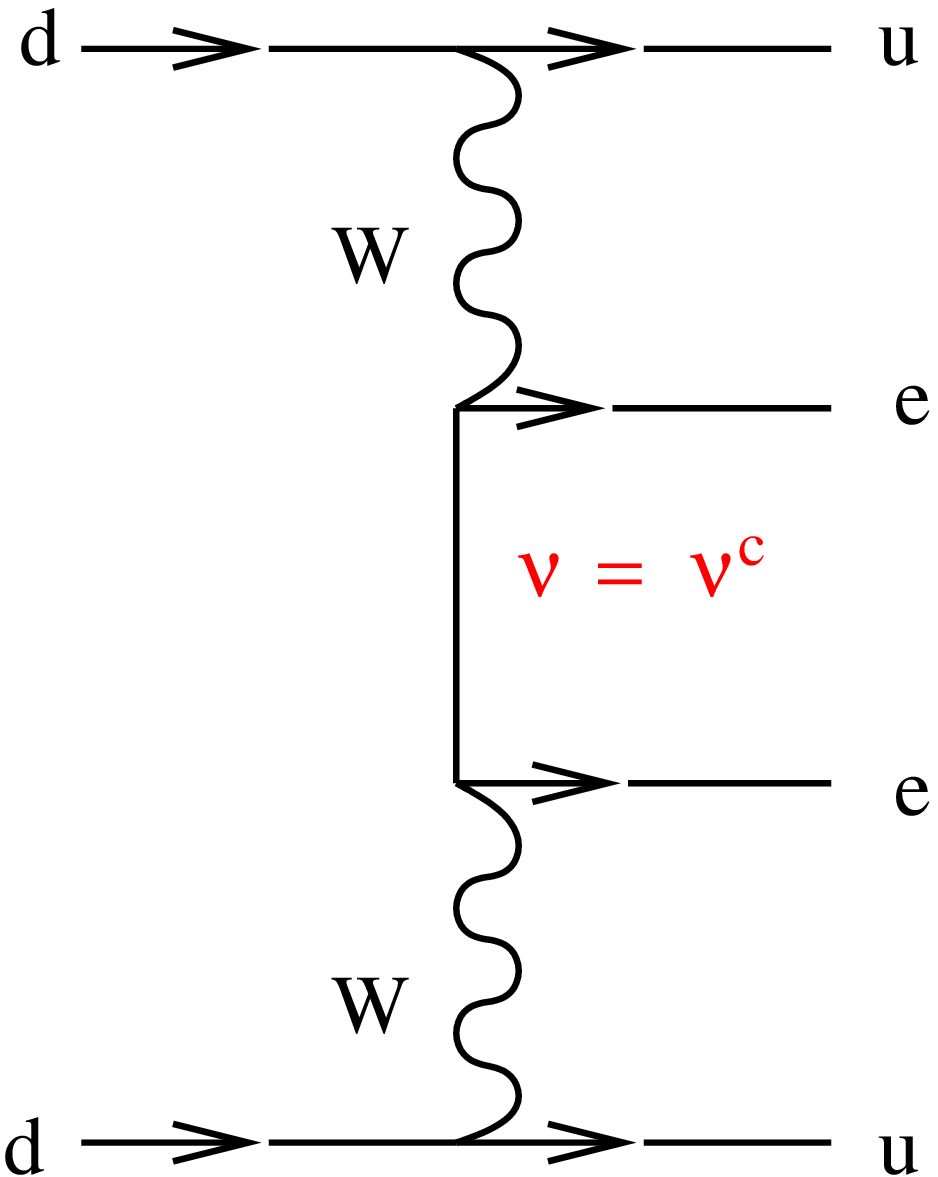} 
\hglue 1cm
\includegraphics[width=5cm,height=3.5cm]{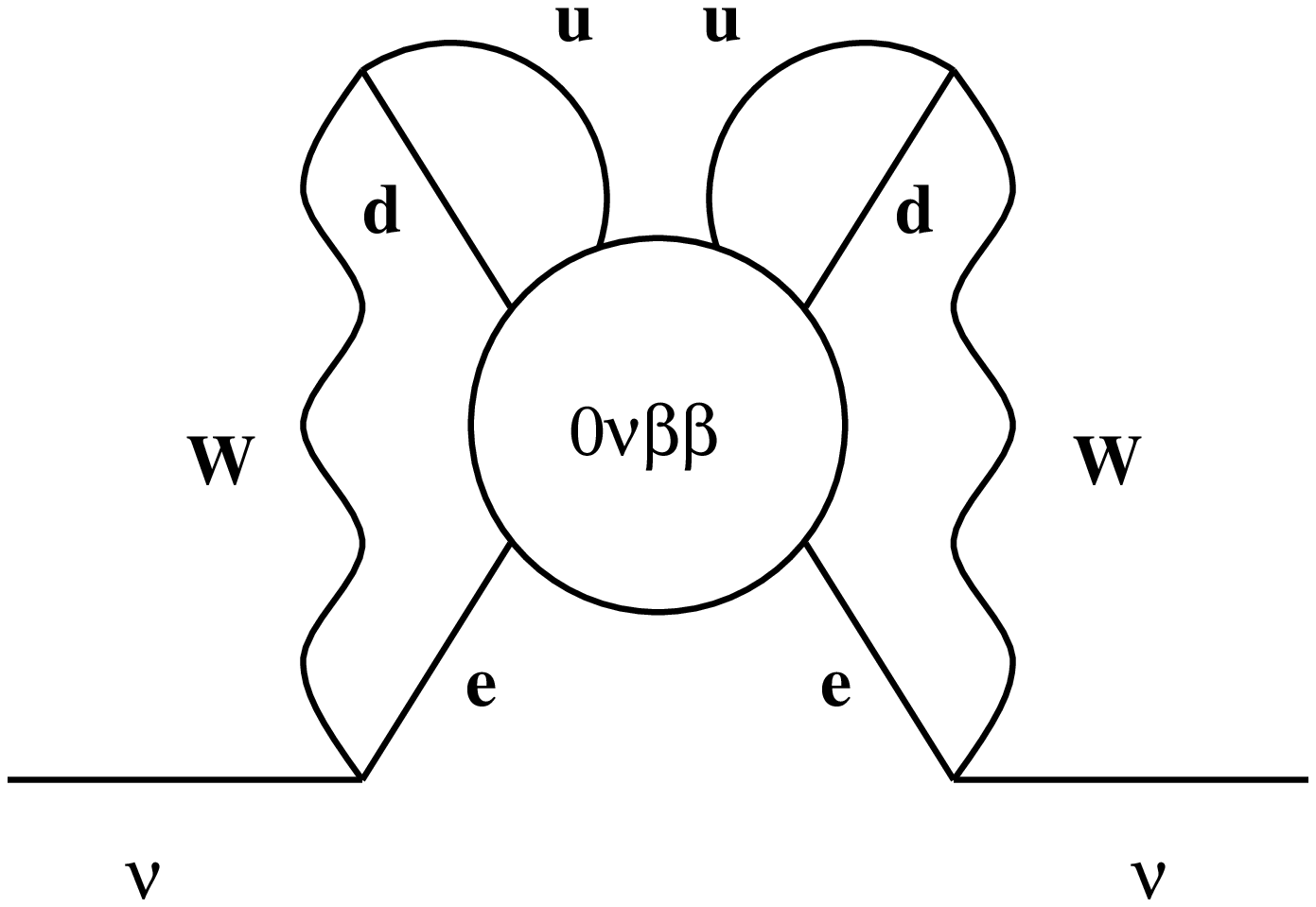}
\caption{Neutrino mass mechanism for \nbb
  (left), and black box theorem (right)~\cite{Schechter:1982bd}.}
 \label{fig:bbox}
\end{figure}

The Dirac or Majorana nature of neutrinos is manifest only through the
observation of LNV processes, such as
\nbb~\cite{schechter:1980gr}~\footnote{Electromagnetic interactions of
  neutrinos can also probe their Majorana nature.}, whose current
status and perspectives was reviewed by
Schoenert~\cite{Avignone:2007fu}.  A nonzero rate for \nbb implies
that, within a gauge theory, at least one neutrino gets a Majorana
mass, this argument is known as the ``black-box''
theorem~\cite{Schechter:1982bd}, illustrated in Fig.
\ref{fig:bbox}, and recently discussed in~\cite{Hirsch:2006yk}.\\
\vskip .1cm
Given the neutrino oscillation data it follows that light Majorana
neutrino exchange will induce \nbb and the corresponding rate is a
measure of the absolute scale of neutrino mass, complementary to the
one probed in beta decay studies~\cite{Drexlin:2005zt}, and
cosmological observations~\cite{Lesgourgues:2006nd}.  It is also
sensitive to the Majorana CP violation~\cite{Schechter:1981gk} which
drops out of oscillations.
Using current neutrino oscillation parameters and state-of-the-art
nuclear matrix elements~\cite{Faessler:2008xj} one can determine the
average mass parameter $\meff$ characterizing the neutrino exchange
contribution to \nbb.

Inverted hierarchy implies a generic lower bound for the \nbb
amplitude, while for normal hierarchy neutrino spectra the three
neutrinos can interfere destructively, so that no generic lower bound
exists. Specific flavor models may provide a lower bound for \nbb even
with normal hierarchy, as discussed
in~\cite{Hirsch:2009mx}~\cite{Hirsch:2005mc,Hirsch:2008rp}.
Quasi-degenerate neutrinos~\cite{babu:2002dz,Hirsch:2003dr} give the
largest possible \nbb signal.  

\vspace{.2cm}

{\em Acknowledgments:}

Work supported by Spanish grants FPA2008-00319/FPA and
PROMETEO/2009/091 and by European Union network UNILHC
(PITN-GA-2009-237920).  \vglue .3cm \renewcommand{\baselinestretch}{1}

\bibliographystyle{iopart-num}   %

\providecommand{\newblock}{}

\end{document}